\documentclass[prl, a4paper,nofootinbib, twocolumn,showpacs,floatfix]{revtex4}

\usepackage{amssymb, amsfonts, amsmath,latexsym, bm, graphicx,verbatim}
\renewcommand{\vec}[1]{{\mathbf #1}}

\def\be{\begin{equation}}
\def\ee{\end{equation}}
\begin{document}
\title{Spin Dynamics in Pyrochlore Heisenberg Antiferromagnets}
\author{P.\ H.\ Conlon}
\date{April 9, 2009}
\email{conlon@thphys.ox.ac.uk}
\author{J.\ T.\ Chalker}
\affiliation{Theoretical Physics, Oxford University, 1, Keble Road, Oxford OX1 3NP, United Kingdom}
\begin{abstract}
We study the low temperature dynamics of the classical Heisenberg antiferromagnet with nearest neighbour interactions on the pyrochlore lattice. We present extensive results for the wavevector and frequency dependence of the dynamical structure factor, obtained from simulations of the precessional dynamics. We also construct a solvable stochastic model for dynamics with conserved magnetisation, which accurately reproduces most features of  the precessional results.
Spin correlations relax at a rate independent of wavevector and proportional to temperature.
\end{abstract}
\maketitle

Geometrical frustration in magnets inhibits ordering. Simple, classical models for these systems have very degenerate ground states \cite{Anderson56,Villain79}. Reflecting this degeneracy, highly frustrated magnetic materials characteristically remain in the paramagnetic phase even at temperatures low compared to the scale set by exchange interactions. Behaviour in this cooperative paramagnetic regime has been the focus of much recent research
\cite{Ramirez94,MoessnerRamirez06}.

Nearest neighbour antiferromagnets on the pyrochlore lattice with classical $n$-component spins are representative of a large class of models \cite{MoessnerChalker98,CanalsGaranin01}. They have remarkable correlations at low temperature, which are intermediate between those of conventionally ordered and completely disordered systems. These can be understood by mapping spin states onto configurations of a vector field, or {\it flux field}, which is solenoidal for ground states  \cite{HuseKrauthMoessnerSondhi03,IsakovGregor04,Henley05}. Gaussian fluctuations of this flux field provide a coarse-grained description of the cooperative paramagnet. Static spin correlations have a power-law dependence on separation, inside a correlation length $\xi$ that diverges as temperature $T$ approaches zero. These correlations result in sharp features, termed {\it pinch points}, in diffuse scattering as a function of wavevector.

The dynamics of cooperative paramagnets has not been studied as extensively as the statics, but some ingredients are clear. In a Heisenberg model with precessional dynamics, the short-time behaviour can be viewed in terms of harmonic spinwave fluctuations in the vicinity of a specific ground state, while over longer times the system wanders around the ground state manifold.  This second component to the motion results in decay of the spin autocorrelation function at long times, with a decay rate shown to be linear in $T$ using simulations and phenomenological arguments \cite{MoessnerChalker98}. These theoretical ideas are supported by inelastic neutron scattering measurements on pyrochlore antiferromagnets: an early study of $\rm Cs Ni Cr F_6$  revealed strong temperature dependence to the width in energy of quasielastic scattering for $T< |\Theta_{\rm CW}|$ \cite{HarrisZinkinZeiske95}, while recent work  on $\rm Y_2Ru_2O_7$ shows a width  linear in $T$ as predicted \cite{vanDuijn08}.

Our aim in this paper is to establish a much more comprehensive description of cooperative paramagnets with precessional dynamics than has been available so far. The topic is interesting from several perspectives. First, in view of the pinch points in static correlations, it is natural to ask about the wavevector dependence of the  dynamical structure factor, accessible in single-crystal measurements.
Little is currently known about this:  the autocorrelation function
of  Ref \cite{MoessnerChalker98} is expressed as an
integral over all wavevectors, while the measurements of
Ref. \cite{vanDuijn08} used a powder sample. Second, dynamics in the
paramagnetic phase of unfrustrated antiferromagnets is dominated by
spin diffusion
\cite{deGennes58,deLeenerResibois66,BunkerChenLandau1996}, and one
would like to know whether this extends to the cooperative
paramagnet. Third, behaviour in the Heisenberg model should be compared to that in spin ice, which is repesented by the Ising pyrochlore antiferromagnet with dynamics controlled by the motion of monopole excitations \cite{Jaubert09}.

In outline, our results are as follows. We find at low temperature three types of behaviour in different regions of reciprocal space. (i) Close to reciprocal lattice points, correlations are dominated by spin diffusion with a temperature-independent diffusion constant. (ii) At a generic wavevector [not included in (i) or (iii)] correlations are Lorentzian in frequency with a width linear in $T$ and independent of wavevector.  (iii)  Close to nodal lines in reciprocal space on which the static, ground-state structure factor vanishes \cite{Henley05}, dynamical correlations are dominated by finite-frequency spinwave contributions. This picture is hence very different from that for the kagome Heisenberg antiferromagnet, which shows order-by-disorder and propagating modes \cite{RobertCanals08}.

We consider the classical Heisenberg antiferromagnet with nearest neighbour interactions on the pyrochlore lattice. Lattice sites (labelled $i,j$) form corner-sharing tetrahedra (labelled $\alpha,\beta$). Spins $\vec{S}_i$ are unit vectors and  $\vec{L}_{\alpha} = \sum_{i\in\alpha}\vec{S}_i$ is the total spin of tetrahedron  $\alpha$. The Hamiltonian is
\be\label{eqn:hamiltonian}
H = J\sum_{\langle ij \rangle} \vec{S}_i.\vec{S}_j \equiv \frac{1}{2}J\sum_{\alpha}\vec{L}_{\alpha}^2 +{\rm  c},
\ee
where $\rm c$ is a constant. Ground states satisfy $\vec{L}_\alpha=0$ for all $\alpha$. The equation of motion, describing  precession of each spin around its local exchange field, is
\be\label{eqn:heisenbergeom}
\frac{d\vec{S}_i}{dt} =  -J \vec{S}_i\times\sum_{j} \vec{S}_j
\ee
where sites $j$ are the nearest neighbours of $i$.
The global spin rotation symmetry of Eq.~(\ref{eqn:hamiltonian}) implies conservation of total spin in the dynamics.

Before presenting results from a molecular dynamics study of
Eq.~(\ref{eqn:heisenbergeom}), we consider an analytically tractable
stochastic model for the dynamic behavior. It is known that static
spin correlators for the classical Heisenberg model are well described
by those for $n$-component spins in the large $n$ limit
\cite{CanalsGaranin01,IsakovGregor04}.  Building on this, we set out
to endow the $n=\infty$ model with appropriate dynamics. First we
recall some details of the static model. Taking the second form
of the Hamiltonian in Eq.~(\ref{eqn:hamiltonian}), a single spin
component in the large $n$ limit has the unnormalised probability
distribution $e^{-\beta E}$ with
\be\label{eqn:partition}
\beta E = \frac{1}{2}\sum_i \lambda s_i^2+ \frac{1}{2}\beta
J\sum_{\alpha}l_{\alpha}^2\;,
\ee
where now $l_{\alpha} = \sum_{i\in\alpha}s_i$ is the sum of `soft'
spins ($-\infty<s_i<\infty$) on tetrahedron $\alpha$. The spin length
is constrained by the Lagrange multiplier $\lambda$. For
$\beta\rightarrow\infty$, the second term in Eq.~(\ref{eqn:partition})
enforces all the $l_{\alpha}$ to be zero. The interaction term written
directly in terms of the spins is $\frac{1}{2}\beta
J\sum_{ij}(A_{ij}+2\delta_{ij})s_is_j$ where $A_{ij}$ is the adjacency
matrix for the pyrochlore lattice. We call the combination
$A_{ij}+2\delta_{ij}$ the interaction matrix. Its eigenvalues
$v_{\mu}(\vec{q})$ are labeled by wavevector $\vec{q}$ and a band
index $\mu \in \{1,2,3,4\}$. Two bands are flat ($v_{1,2}(\vec{q})=0$)
and two ($\mu=3,4$) are dispersive. Requiring $\langle s_i^2 \rangle
=1/3$ to mimic behaviour of a single spin component in the Heisenberg
model, $\lambda = 3/2 +{\cal O}(T/J)$ for $T \ll J$. We denote the
Fourier transform of the spin variables $s_i$ by $s^{a}_{\vec{q}}$,
where $a$ is a sublattice index, and define the sublattice sum
$s_{\vec{q}} = \sum_{a=1}^4 s^a_{\vec{q}}$. Transforming from $s^a_{\vec{q}}$ to the basis
(denoted by tildes) that diagonalises the interaction matrix  gives
collective spin variables $\tilde s^{\mu}_{\vec{q}}$. We want to
introduce time dependence and calculate the dynamic correlation
function $S(\vec{q},t)=\langle s_{\vec{q}}(t)s^*_{\vec{q}}(0)\rangle$,
and its time
Fourier transform, the dynamic structure factor $S(\vec{q},\omega)$,
measured using neutron scattering.

There are many choices of dynamics which reproduce any given
equilibrium distribution. To approximate Eq.~(\ref{eqn:heisenbergeom})
we demand a local dynamics that conserves the total spin. We can
ensure this by requiring the spin on each site to satisfy a local continuity equation. We
introduce spin currents on bonds of the pyrochlore lattice, which have drift and noise terms.
We take the drift current on a bond linking two site to be  proportional to the difference in the generalized forces $\partial E/\partial s_i$ at the sites. This favours relaxation towards a configuration that minimizes $E$; the thermal ensemble is maintained by noise which has an independent Gaussian distribution on each bond. These assumptions lead to the dynamical equations for the soft spins
\be \label{eqn:langevineom}
\frac{d s_i}{dt} = \Gamma \sum_{l}\Delta_{il} \frac{\partial E}{\partial s_l} + \zeta_i(t)
\ee
where the matrix $\Delta$ is the lattice laplacian (for a lattice with coordination number $z$, $\Delta_{il} = A_{il} - z \delta_{il}$). The correlator of the noise $\zeta_i(t)$  at site $i$, $\langle\zeta_i(t)\zeta_j(t')\rangle=2T\Gamma\Delta_{ij}\delta(t-t')$, has an amplitude fixed by the requirement of thermal equilibrium. The only free parameter in the model is the rate $\Gamma$, which sets a timescale for dynamical processes. The Langevin equation Eq.~(\ref{eqn:langevineom}) is straightforward to solve in the diagonal basis. It gives the correlation function
\be\label{eqn:diagonal}
\langle \tilde s^{\mu}_{\vec{q}}(t) \tilde s^{\nu}_{-\vec{q}}(0) \rangle = \frac{\delta_{\mu\nu}T}{Jv_{\mu}+\lambda T}e^{-\Gamma(8-v_{\mu})(Jv_{\mu}+\lambda T)t}\;.
\ee
The dynamic correlation function is then
\be \label{eqn:dynstructure}
S(\vec{q},t)
= \sum_{\mu=1}^4g_{\mu}(\vec{q})\langle \tilde s^{\mu}_{\vec{q}}(t) \tilde s^{\mu}_{-\vec{q}}(0) \rangle
\ee
where the structure factors $g_{\mu}(\vec{q})$ are formed from the eigenvectors of the interaction matrix. They satisfy the sum rule $\sum_{\mu=1}^4 g_{\mu}(\vec{q})=4$.

For completeness we present their explicit forms here. In the notation of \cite{IsakovGregor04}, where $c_{ab} = \cos\left(\frac{q_a+q_b}{4}\right)$ and $c_{\overline{ab}} = \cos\left(\frac{q_a-q_b}{4}\right)$ with
$Q = c_{xy}^2 +c_{\overline{xy}}^2+c_{yz}^2+c_{\overline{yz}}^2+c_{xz}^2+c_{\overline{xz}}^2-3$ and defining $P \equiv \sqrt{1+Q}$, the eigenvalues of the interaction matrix are $v_{1,2} = 0$ and $v_{3,4} = 4 \mp 2P$. Further defining $s^2_{a} \equiv \sin^2\left(\frac{q_a}{4}\right)$ and $c_{(ab)}\equiv c_{ab}+c_{\overline{ab}}$, the $g_{\mu}(\vec{q})$ are for the degenerate flat bands
\begin{equation*}
g\equiv g_1+g_2  = 2 -\frac{4}{3-Q}\biggl[
      c_{(xy)}s^2_z+ c_{(yz)}s^2_x + c_{(zx)}s^2_y
                           \biggr]
\end{equation*}
and for the dispersive bands
\begin{multline*}\label{eqn:gs}
g_{3,4} = 2 - \frac{1}{2}g\left(1\pm 2P^{-1}\right)\pm P^{-1}(2-c_{(yz)}-c_{(xz)}-c_{(xy)})
\end{multline*}
which indeed satisfy $g_1+g_2+g_3+g_4=4$.

We now examine the implications of this model for $T \ll J$,
emphasizing the features (i) -- (iii) mentioned in our introduction.
From the exponent in Eq.~(\ref{eqn:diagonal}) we obtain a
characteristic time for decay of correlations. (i)~In the vicinity of
$\vec{q}=0$ only the coefficient $g_4(\vec{q})$ is non-zero and so behaviour is controlled by the fourth band whose decay rate is
$\tau^{-1} = 8\Gamma Ja^2q^2+{\cal O}(q^4)$, where $a$ is the
pyrochlore site spacing; from this we identify the spin diffusion
constant in this model as $D = 8 \Gamma J a^2$, independent of
$T$. (ii) At a generic wavevector  where $g_1$ and $g_2$ are non-zero
most of the spectral weight is in the flat bands, with decay rate
$\tau^{-1} = 8\Gamma\lambda T$, independent of $q$; in an
approximation where only the flat bands contribute, this implies the
dynamic structure factor factorizes as $S(\vec{q},\omega) =
S(\vec{q})f(\omega)$, a possibility noted in
\cite{CanalsGaranin01}. (iii) On nodal lines \cite{Henley05}, high
symmetry directions in reciprocal space along which
$g_1(\vec{q})+g_2(\vec{q})=0$,  the decay rate is wavevector-dependent and ${\cal O}(J)$ away from $\vec{q}=0$.

\begin{figure}
\includegraphics[width = 83mm]{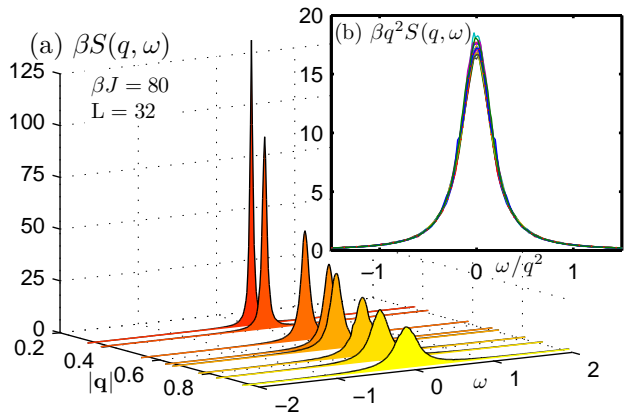}
\caption{(Colour online) Evidence for spin diffusion. (a) Dependence of $S(q,\omega)$ on $q$ and $\omega$ at small $q$. (b) Scaling collapse following Eq.~(\ref{eqn:diffusion}) at multiple temperatures ($\beta J = 20,40,60,80$) and four of the wavevectors  plotted in (a). Also plotted is the prediction of Eq.~(\ref{eqn:dynstructure}).}
\label{fig:diffusion}
\end{figure}
\begin{figure}[t]
\includegraphics[width = 83mm]{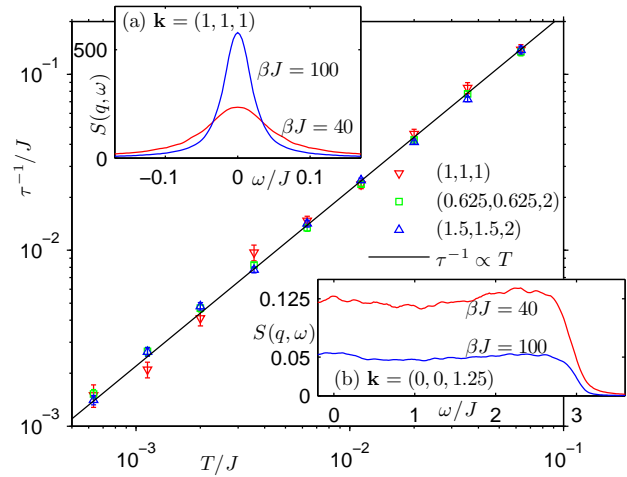}
\caption{(Colour online) Main panel: decay rate of $S(\vec{q},t)$ as a
  function of $T$ at wavevectors as indicated. Upper inset:
  $S(\vec{q},\omega)$ at $\beta J$ = 40 (red) and 100 (blue), for
  $\vec{k}$ = (1,1,1). Lower
  inset:  $S(\vec{q},\omega)$ on a nodal line at
  $\vec{k}=(0,0,1.25)$.}
\label{fig:decaywidth}
\end{figure}
\begin{figure}[t]
\includegraphics[width = 83mm]{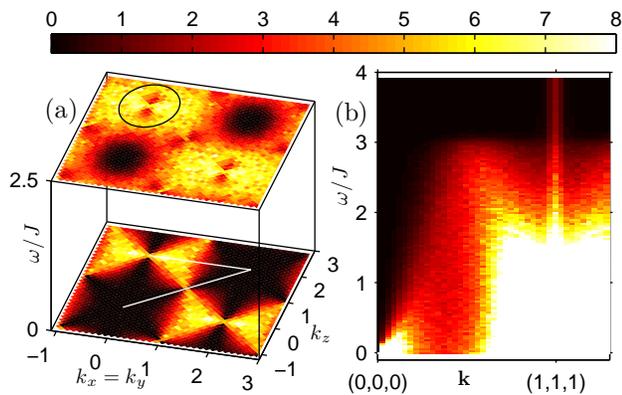}
\caption{(Colour online) Intensity map of $\beta S(\vec{q},\omega)$ in the $(q_x,q_x,q_z)$ plane at $\beta J = 500$. (a) Lower panel: zero frequency (data divided by $1.25\times10^5$); white line is path ${\cal P}$. Upper panel: $\omega=2.5J$; black circle is centered on a pinch-point. (b) Section along path segment  ${\cal P}_1$ (see text).}
\label{fig:sqwplane}
\end{figure}
\begin{figure}[b]
\includegraphics[width = 83mm]{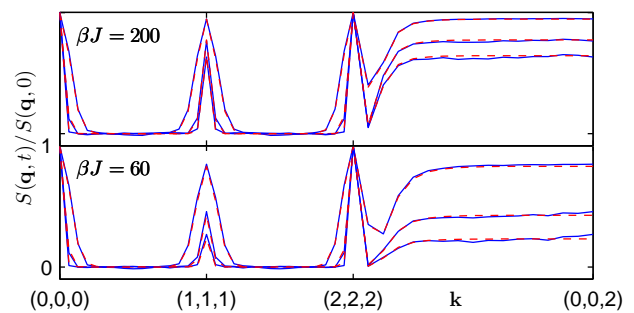}
\caption{(Colour online) Solid blue lines: normalized correlation
  function $S(\vec{q},t)/S(\vec{q},0)$. Dashed red lines: prediction
  from Eq.~(\ref{eqn:dynstructure}) with $\Gamma = 0.167$, both shown at 3 times, $t$.}
\label{fig:sqt}
\end{figure}

To test these ideas we have performed simulations of the full precessional dynamics, Eq.~(\ref{eqn:heisenbergeom}).
Low $T$ configurations are obtained using a Metropolis Monte Carlo
sampling method. We take these as initial configurations for numerical
integration of the equations of motion using a 4th order Runge-Kutta
algorithm with adaptive step size. Energy and total spin are conserved
to relative errors no greater than $10^{-6}$. We report data from
simulations on system sizes with total number of sites $N = 4L^3$ for
$L= 16,32$. We calculate the dynamic correlation function $S(\vec{q},t)\equiv \langle
\vec{S}_{\vec{q}}(t)\cdot\vec{S}_{-\vec{q}}(0)\rangle$ and the dynamical
structure factor $S(\vec{q},\omega) \equiv \langle|
\vec{S}_{\vec{q}}(\omega)|^2\rangle$ where $\vec{S}_{\vec{q}} =
\sum_{a=1}^4\vec{S}^{a}_{\vec{q}}$ is the Fourier transformed spin
configuration and $a$ runs over the four sublattices. We present
results under the headings (i) -- (iii) as above, expressing $\vec{q}$ in reciprocal lattice units
as $\vec{q} = 2\pi\vec{k}$ for Figs.~\ref{fig:decaywidth}, \ref{fig:sqwplane}, \ref{fig:sqt}.

(i) Since the total magnetization is conserved, one expects diffusion at sufficiently small $q$. The simulations confirm diffusive behavior with a diffusion constant independent of temperature. At small $q$, the data should collapse onto the scaling form appropriate for diffusion,
\be\label{eqn:diffusion}
\beta q^2 S(\vec{q},\omega) = 3\chi \frac{2 D}{(\omega/q^2)^2+D^2}\;,
\ee
where $\chi$ is the susceptibility per primitive unit cell and the factor of 3 is due to the 3 spin components. Fig.~\ref{fig:diffusion} shows this scaling collapse when
plotted as in Eq.~(\ref{eqn:diffusion}), and demonstrates that the
diffusion constant is independent of temperature. The prediction of
Eq.~(\ref{eqn:dynstructure}), whose small $q$ limit has precisely the
form of Eq.~(\ref{eqn:diffusion}), is an excellent fit with
$\Gamma = 0.167$.

(ii) At a generic point in reciprocal space (not near $\vec{q}=0$ or a
nodal line) the structure factor is well described by a Lorentzian
centered on $\omega = 0$, indicating relaxational dynamics
(Fig.~\ref{fig:decaywidth}, upper inset). The decay rate for this relaxation
(Fig.~\ref{fig:decaywidth}, main panel) is
proportional to $T$ and independent of wavevector, even close to the pinch points.

(iii) On nodal lines, by contrast, the width in frequency of
$S(\vec{q},\omega)$ is ${\cal O}(J)$ and depends little on $T$: see
Fig.~\ref{fig:decaywidth}, lower inset. High frequency ($\omega=2.5J$)
and zero frequency behaviour is also presented in
Fig.~\ref{fig:sqwplane}, as a survey of $S(\vec{q},\omega)$ in the
$(q_x,q_x,q_z)$ plane, using data
taken at $\beta J = 500$ and typical of all low temperatures.
Weight in $S(\vec{q},\omega)$ at $\omega \sim J$ and low $T$ can be
viewed as due to spinwave fluctuations in the vicinity of an
instantaneous ground state. In contrast to behaviour in the kagome
antiferromagnet \cite{RobertCanals08}, there in no evidence in
Fig.~~\ref{fig:sqwplane} for sharp propagating modes.

We next demonstrate that the stochastic model accurately reproduces
most aspects of the precssional dynamics.
We examine behaviour with each type of dynamics at different temperatures and times
along a path in the Brillouin zone:
${\cal P} = (0,0,0)\overset{{\cal
    P}_1}{\rightarrow}(2,2,2)\overset{{\cal
    P}_2}{\rightarrow}(0,0,2)$. This consists of the section ${\cal P}_1$ along
    high symmetry nodal lines passing through
a pinch-point, and a section ${\cal P}_2$ typical of reciprocal
space. We compare in Fig.~\ref{fig:sqt} the dynamic correlation
function, normalized to unity at $t=0$, at two temperatures with the
predictions from Eq.~(\ref{eqn:dynstructure}) along the path ${\cal
  P}$. The excellent agreement of the curves across
multiple temperatures, wavevectors and times is good evidence that the
stochastic model is sufficient to capture the relaxation behavior. It
fails only at short times ($t\lesssim J^{-1}$, not shown in
Fig.~\ref{fig:sqt}) on nodal lines, where it does not account for
oscillatory spinwave contributions.

The stochastic model is microscopic and its main ingredients are a
conservation law and the pyrochlore lattice structure. A long
wavelength description is provided by the mapping to flux fields
\cite{Henley05,IsakovGregor04} and it is interesting to see how the
dynamics translates under this mapping. Taking the low temperature,
small $q$ limit, the correlators for the continuum flux fields
$\vec{B}(\vec{q},t)$ implied
by the stochastic model are
\begin{multline*}
\langle B_i(\vec{q},t)B_j(-\vec{q},0)\rangle\propto\left(\delta_{ij} - \frac{q_iq_j}{q^2}\right)e^{-8\Gamma\lambda Tt}\\
+\left(\frac{q_iq_j}{q^2}-\frac{q_iq_j}{q^2+\xi^{-2}}\right)e^{-8\Gamma(Ja^2q^2+\lambda T)t}\;.
\end{multline*}
This result can be derived from a Langevin equation for the continuum
flux fields in which the `monopole density' $\rho = \nabla \cdot
\vec{B}$ obeys a continuity equation $\partial_t \rho + \nabla \cdot
\vec{j} =0$, with monopole current density
\be\label{eqn:continuumcurrent}
\vec{j} = 8\Gamma\lambda T \vec{B} - 8\Gamma J a^2\nabla \rho +\vec{\eta}(t)\;.
\ee
Here, the second term is the usual diffusion current arising from a
density gradient, while the first describes response to an entropic
force. This response involves a drift current of the magnetic charge
density $\rho$ in the field $\vec{B}$ that mimics electrical
conduction in electrodynamics and is responsible for the flat
relaxation rate. Related results have been obtained recently in a
study of dynamics in spin ice, represented by the Ising
antiferromagnet \cite{Jaubert09}. In this case monopoles are discrete
and it has been argued that purely diffusive dynamics are insufficient
to explain observations and a full description must
include the network of Dirac strings between monopoles, which are
essentially entropic, as well as dipolar interactions \cite{Jaubert09}. In spin ice, dipolar interactions lead to Coulomb-law forces between monopoles. By contrast, in Eq.~\ref{eqn:continuumcurrent} Coulombic forces appear purely entropically.

In summary, we have considered wavevector and frequency resolved
dynamics of the classical pyrochlore antiferromagnet. The relaxational
behavior is well captured by a stochastic model that conserves total
spin. Spin diffuses with a diffusion constant independent of
temperature, and entropic forces drive currents to relax
configurations with a rate independent of wave vector and inversely
proportional to temperature.

\begin{acknowledgments}
This work was supported in part by EPSRC Grant No.
EP/D050952/1.
\end{acknowledgments}

\begin{thebibliography}{17}
\providecommand{\bibinfo}[2]{#2}
\providecommand{\eprint}[2][]{\url{#2}}

\bibitem{Anderson56}
\bibinfo{author}{{P.~W.} {Anderson}},
  \bibinfo{journal}{Phys. Rev.} \textbf{\bibinfo{volume}{102}},
  \bibinfo{pages}{1008} (\bibinfo{year}{1956}).

\bibitem{Villain79}
\bibinfo{author}{{J.}~{Villain}}, \bibinfo{journal}{Z.
  Phys. B} \textbf{\bibinfo{volume}{33}}, \bibinfo{pages}{31}
  (\bibinfo{year}{1979}).

\bibitem{Ramirez94}
\bibinfo{author}{{A.~P.} {Ramirez}},
  \bibinfo{journal}{Annu. Rev. Mater. Sci.} \textbf{\bibinfo{volume}{24}},
  \bibinfo{pages}{453} (\bibinfo{year}{1994}).

\bibitem{MoessnerRamirez06}
\bibinfo{author}{{R.}~{Moessner}} {and}
  \bibinfo{author}{{A.~P.} {Ramirez}},
  \bibinfo{journal}{Physics Today} \textbf{\bibinfo{volume}{59}},
  \bibinfo{pages}{24} (\bibinfo{year}{2006}).

\bibitem{MoessnerChalker98}
\bibinfo{author}{{R.}~{Moessner}} {and}
  \bibinfo{author}{{J.~T.} {Chalker}},
  \bibinfo{journal}{Phys. Rev. B} \textbf{\bibinfo{volume}{58}},
  \bibinfo{pages}{12049} (\bibinfo{year}{1998}{\natexlab{a}});
  \bibinfo{journal}{Phys. Rev. Lett} \textbf{\bibinfo{volume}{80}},
  \bibinfo{pages}{2929} (\bibinfo{year}{1998}{\natexlab{b}}).

\bibitem{CanalsGaranin01}
\bibinfo{author}{{B.}~{Canals}} {and}
  \bibinfo{author}{{D.~A.} {Garanin}},
  \bibinfo{journal}{Can. J. Phys} \textbf{\bibinfo{volume}{79}},
  \bibinfo{pages}{1323} (\bibinfo{year}{2001}).

\bibitem{IsakovGregor04}
\bibinfo{author}{{S.~V.} {Isakov}},
  \bibinfo{author}{{K.}~{Gregor}},
  \bibinfo{author}{{R.}~{Moessner}}, {and}
  \bibinfo{author}{{S.~L.} {Sondhi}},
  \bibinfo{journal}{Phys. Rev. Lett.} \textbf{\bibinfo{volume}{93}},
  \bibinfo{pages}{167204} (\bibinfo{year}{2004}).

\bibitem{Henley05}
\bibinfo{author}{{C.~L.} {Henley}},
  \bibinfo{journal}{Phys. Rev. B} \textbf{\bibinfo{volume}{71}},
  \bibinfo{eid}{014424} (\bibinfo{year}{2005}).

\bibitem{HuseKrauthMoessnerSondhi03}
\bibinfo{author}{{D.~A.} {Huse}},
  \bibinfo{author}{{W.}~{Krauth}},
  \bibinfo{author}{{R.}~{Moessner}}, {and}
  \bibinfo{author}{{S.~L.} {Sondhi}},
  \bibinfo{journal}{Phys. Rev. Lett} \textbf{\bibinfo{volume}{91}},
  \bibinfo{pages}{167004} (\bibinfo{year}{2003}).

\bibitem{HarrisZinkinZeiske95}
\bibinfo{author}{{M.~J.} {Harris}},
  \bibinfo{author}{{M.~P.} {Zinkin}},
  {and} \bibinfo{author}{{T.}~{Zeiske}},
  \bibinfo{journal}{\prb} \textbf{\bibinfo{volume}{52}},
  \bibinfo{pages}{R707} (\bibinfo{year}{1995}).

\bibitem{vanDuijn08}
\bibinfo{author}{{J.}~{van Duijn}},
  \bibinfo{author}{{N.}~{Hur}},
  \bibinfo{author}{{J.~W.} {Taylor}},
  \bibinfo{author}{{Y.}~{Qiu}},
  \bibinfo{author}{{Q.~Z.} {Huang}},
  \bibinfo{author}{{S.-W.} {Cheong}},
  \bibinfo{author}{{C.}~{Broholm}}, {and}
  \bibinfo{author}{{T.~G.} {Perring}},
  \bibinfo{journal}{\prb} \textbf{\bibinfo{volume}{77}}, \bibinfo{eid}{020405(R)}
  (\bibinfo{year}{2008}).

\bibitem{deGennes58}
\bibinfo{author}{{P.~G.} {de~Gennes}},
  \bibinfo{journal}{J. Phys. Chem. Solids} \textbf{\bibinfo{volume}{4}},
  \bibinfo{pages}{223} (\bibinfo{year}{1958}).

\bibitem{deLeenerResibois66}
\bibinfo{author}{{M.}~{de~Leener}} {and}
  \bibinfo{author}{{P.}~{R\'esibois}},
  \bibinfo{journal}{Phys. Rev.} \textbf{\bibinfo{volume}{152}},
  \bibinfo{pages}{318} (\bibinfo{year}{1966}).

\bibitem{BunkerChenLandau1996}
\bibinfo{author}{{A.}~{Bunker}},
  \bibinfo{author}{{K.}~{Chen}}, {and}
  \bibinfo{author}{{D.~P.} {Landau}},
  \bibinfo{journal}{Phys. Rev. B} \textbf{\bibinfo{volume}{54}},
  \bibinfo{pages}{9259} (\bibinfo{year}{1996}).

\bibitem{Jaubert09}
\bibinfo{author}{{L.}~{Jaubert}} {and}
  \bibinfo{author}{{P.}~{Holdsworth}},
  \bibinfo{journal}{Nature Physics} \textbf{\bibinfo{volume}{5}}, 258 (\bibinfo{year}{2009}).

\bibitem{RobertCanals08}
\bibinfo{author}{{J.}~{Robert}},
  \bibinfo{author}{{B.}~{Canals}},
  \bibinfo{author}{{V.}~{Simonet}}, {and}
  \bibinfo{author}{{R.}~{Ballou}},
  \bibinfo{journal}{Phys. Rev. Lett.} \textbf{\bibinfo{volume}{101}},
  \bibinfo{eid}{117207} (\bibinfo{year}{2008}),

\end{thebibliography}

\end{document}